\documentclass[5p,authoryear,preprint,times]{elsarticle}

\usepackage{color}
\usepackage[]{graphicx}
\usepackage{latexsym}
\usepackage{natbib}
\usepackage{amsmath}
\usepackage{amsbsy}
\usepackage[caption=false]{subfig}
\usepackage{multirow}
\usepackage{mathtools}
\usepackage{textcomp}
\usepackage{rotating}
\usepackage{tikz}

\usepackage[pdftex,hyperfigures,breaklinks,colorlinks,citecolor=black,linkcolor=black]{hyperref} 
\setcitestyle{author,authoryear,open={(},close={)},semicolon}
\usepackage{sectsty}
\allsectionsfont{\centering}

\usepackage{xpatch}
\xpatchcmd{\MaketitleBox}{\hrule}{}{}{}
\xpatchcmd{\MaketitleBox}{\hrule}{}{}{}

\usepackage{etoolbox}
\patchcmd{\abstract}{Abstract}{}{}{}

\begin{document}

\title{How the Brain Transitions from Conscious to Subliminal
  Perception}

\author[1]{Francesca Arese Lucini}
\author[1,2]{Gino Del Ferraro}
\author[3,4,5]{Mariano Sigman}
\author[1]{Hern\'{a}n A. Makse}

\address[1]{Levich Institute and Physics Department, City College of
  New York, New York, NY 10031, USA}
  \address[2]{Department of Radiology, Memorial Sloan Kettering Cancer Center, New York, NY 10065, USA} 
\address[3]{Laboratorio de Neurociencia, Universidad Torcuato Di Tella, Av. Pres. Figueroa Alcorta 7350, Buenos Aires, Argentina}
\address[4]{CONICET (Consejo Nacional de Investigaciones Científicas y Técnicas), Sarmiento 440, Buenos Aires, Argentina}
\address[5]{Facultad de Lenguas y Educaci\'{o}n, Universidad Nebrija,Calle de Sta. Cruz de Marcenado, 27, 28015 Madrid, Spain}

\begin{abstract}  

  \textbf{Abstract--We study the transition in the functional networks that characterize
  the human brains' conscious-state to an unconscious subliminal state
  of perception by using $k$-core percolation. We find that the most
  inner core (i.e., the most connected kernel) of the conscious-state
  functional network corresponds to areas which remain functionally active
  when the brain transitions from the conscious-state to the
  subliminal-state. That is, the inner core of the conscious network
  coincides with the subliminal-state.  Mathematical modeling allows
  to interpret the conscious to subliminal transition as driven by
  $k$-core percolation, through which the conscious state is lost by
  the inactivation of the peripheral $k$-shells of the conscious
  functional network. Thus, the inner core and most robust component
  of the conscious brain corresponds to the unconscious subliminal
  state. This finding imposes constraints to theoretical models of
  consciousness, in that the location of the core of the functional
  brain network is in the unconscious part of the brain rather than in
  the conscious state as previously thought.} \\
  

\end{abstract}

\maketitle

\section*{Introduction}\label{sec:intro}

The human brain as a natural system has received growing attention. 
The scientific literature has explored, from a mathematical and theoretical physics
perspective, the sensitivity and relevance of different properties of
brain topology from a network standpoint ~\citep{framework,greicius_cij,sporns_connectome,sporns_kcore,vandenheuvel_cij,sporns_AM,gallos_percolation,craddock_cij,sporns2013,deco_functional}. 
Many scales and levels of
detail have been investigated, from completely defined networks of hundreds of neurons in
species with particularly small brains \citep{celegans} to macroscopic
summaries of networks up to 100 billion neurons of the mammalian brain
\citep{sporns_connectome,sporns2013,gino_LTP,monkeys}.

Earlier studies concentrated on the
distribution of degree (the number of neighbors on each node), the
clustering (the likelihood that co-neighbors of a node will also be
neighbors), or the diameter (the typical distance between two nodes of
the network) \citep{sporns_connectome,net_measures,sporns_AM,bardella2016hierarchical}. A
second wave of studies has
combined these measures together, for instance in the notion of
small-world networks and weak links~\citep{watts-strogatz,gallos_percolation}. 

Several other statistical
markers of networks have been investigated and recently the idea of
\textit{$k$-core} \citep{intro_kcore,wormald} has received
substantial attention in network analysis since it 
provides a topological notion of the structural skeleton of a network
\citep{dorogovtsev_kcore,internet_kcore,kdecomp,sporns_kcore,kitsak}. 
Theoretical analyses \citep{kcore} have shown that, the $k$-core  may also be an 
indicator for the stability of complex biological systems.
Of particular importance for the scope of this paper is the analysis done in ~\citep{sporns_kcore,shlomo}, 
which demonstrate that the k-core of the network is located in the posterior regions of the brain.

In this work we use network measures as a tool to inquire on one of the most challenging questions in
brain science: the signatures of conscious and subliminal perception. 
We use the notion of $k$-core derived from theoretical physics,
as a fundamental measure of centrality and robustness within a
network, to address the question arising from brain science concerning what brain markers
characterize the conscious $\to$ subliminal transition. 

Recent theoretical results \citep{kcore} highlight how the resilience of neural dynamical systems is controlled by the strength of the interaction couplings and that, furthermore, the most robust part of the system under interaction coupling change is the maximum $k$-core of the corresponding network. This study inspired us to investigate the maximum $k$-core of the network for the system under study and verify if it has a neuroanatomical correspondence. If one can give meaning to such robust subset of the network, then the $k$-core percolation would represent a meaningful method to model of the corresponding transition. 

We build on a classic
study of human brain activations performed by Dehaene \textit{et al.} \citep{experiment}. These experiments measure, through functional Magnetic
Resonance Imaging (fMRI), participants that either record seeing an image flashed at millisecond intervals on a computer screen in front of them (conscious state), or they do not (subliminal state). We build the functional brain networks of the obtained conscious
state based on temporal similarity of activations. 

The main
theoretical question that we then ask is how the transition from conscious to subliminal
state can be modeled in terms of network theory and what subset of the conscious-state network describes the final state of this transition to the subliminal state.
We contrast two possible hypotheses. A natural idea is that the
$k$-core decomposition method may index regions that are more
relevant for conscious processing. This intuition comes from several
theoretical studies of the neural substrate of
consciousness advanced by Dehaene, Tononi and collaborators ~\citep{book,tononi_consc}, which argue that vast
broadcasting, dense and flexible connectivity may be a central
feature of consciousness.  Differently, several psychological
theories, most notably deriving from the work of Benjamin Libet
~\citep{Libet}, have implied that subliminal processing provides a
kernel for all thought. In this view, consciousness is `merely' a
read-out of a vast and robust cascade of processes. 

Discriminating between these two theories requires 
to understand whether the $k$-core of a set of
conscious activations is associated with specific nodes of the
network that make this activation conscious or, instead, with a
subliminal stream which serves as a structural core for subsequent
conscious activations. Our analysis supports the latter hypothesis:
the functional network which models the subliminal-state of the brain corresponds to 
the maximum $k$-core of the more extended functional network which models the conscious-state.

The article is organized as follows. First we give an illustration of
the experiments performed by Dehaene \textit{et al}. \citep{experiment}, of which we analyze 
all data, and we then provide 
a corresponding definition of conscious- and subliminal-state of the brain. 
In Section Experimental Procedure we describe the methodology employed to 
construct functional brain networks of  the conscious-state and introduce the 
concept of $k$-core decomposition as a trimming process to identify network structures. 
Section Results discusses our findings and shows that nodes in the maximum $k$-core 
of the conscious-state network correspond to the subliminal-state of fMRI activation, both at the brain
module- and node-level. In Section Discussion we elaborate on the interpretation of our results by summarizing the theory developed in \cite{kcore} on the role of the $k$-core as indicator of network robustness. In the same section we also contextualize our findings within two consciousness theories developed by Libet and Dehaene-Tononi. Section Conclusions summarizes the study and draws the conclusions.

\section*{Data} \label{sec:data}

\begin{figure}[t!]
\centering
\includegraphics[width=0.5\textwidth]{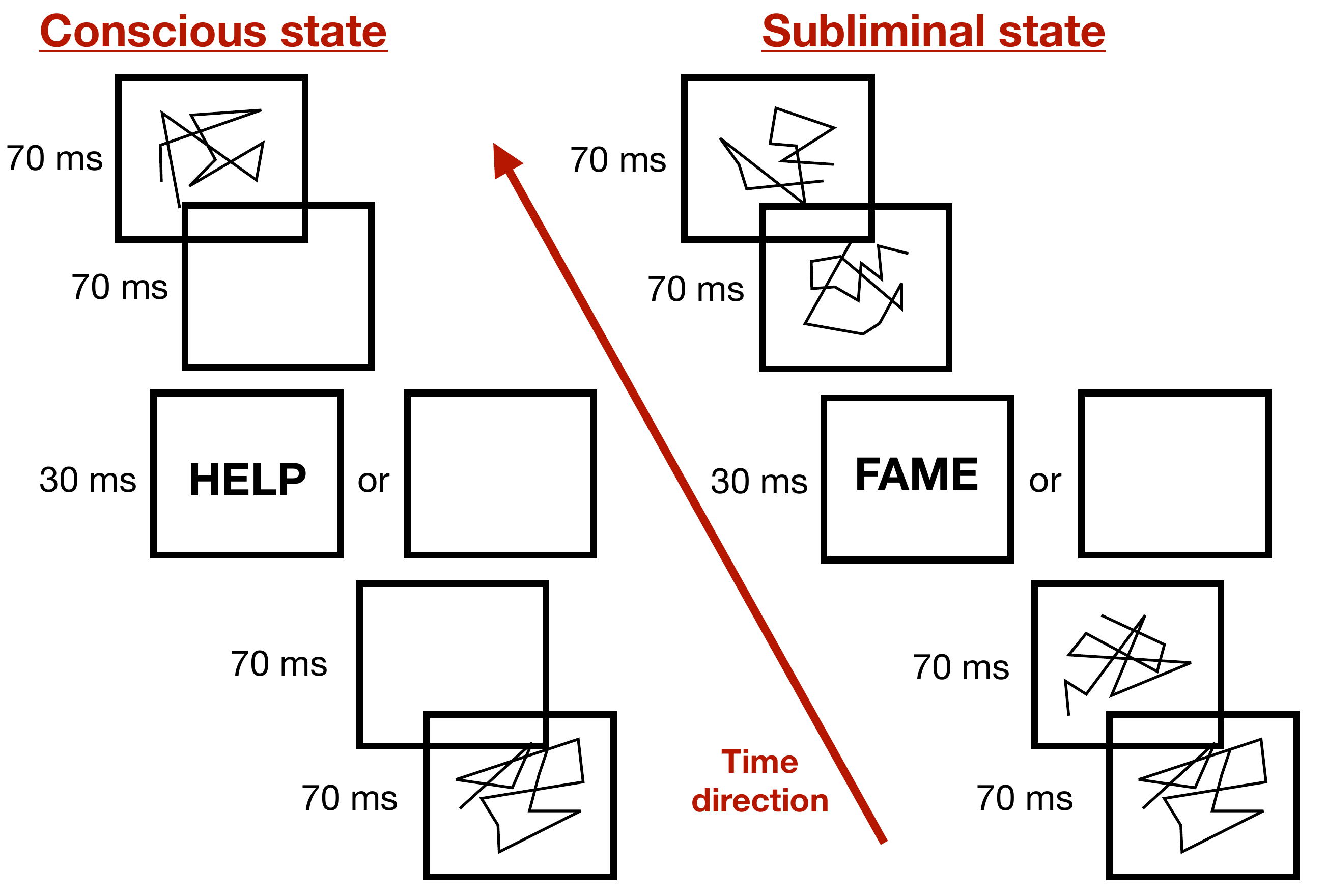}
\caption{ {\bf{Simplified sketch of the experiment}} \citep{experiment}. The left illustration portrays the stream sequence used to cause the conscious-state perception, where the four lettered word is presented preceded and succeeded by blank screens. The right illustration portrays the experiment where the word is sandwiched between distractors, or masks, which inhibits the conscious perception of the word and causes the subliminal-state activation. For each of these two experiments a control sequence is presented in which blank images are displayed instead of words.
}
\label{experiment}
\end{figure}

The data we use in our study and analysis were collected by ~\cite{experiment} and are briefly
explained next.
In the investigation discussed in~\cite{experiment} a subject endures two different experiments for a specific time interval. In this time frame 
four letter words are presented to a participant who undergoes fMRI screening. Each word is flashed on a computer screen either sandwiched between blank pictures or preceded and succeeded by images on the screen called distractors or
masks~\citep{experiment}, as illustrated in Fig. \ref{experiment}. 
Words in both scenarios are flashed for 
30ms and the sequence of blank screens and words (or masks) is repeated, with a fixed order, for a total of 5 minutes. 

In the first type of experiments a word is flashed on the computer screen sandwiched between blank images, designed to produce a {\it conscious} perception of the word by the subject, who, indeed, reports to have seen the word on the screen after each stream of images. We will refer to the fMRI signal of this state as {\it conscious} or {\it unmasked} (see Fig. \ref{experiment}). The second type of experiments are, on the contrary, designed not to produce any active perception of the word, which is in fact flashed sandwiched between scrambled words. The distractor images, indeed, act as `masks' and the subject does not consciously detect the word on the screen. We will refer to the fMRI signal of this state as {\it masked} or {\it subliminal} (see Fig. \ref{experiment}).
\begin{figure*}[!ht]
\centering
\includegraphics[width=0.8\textwidth]{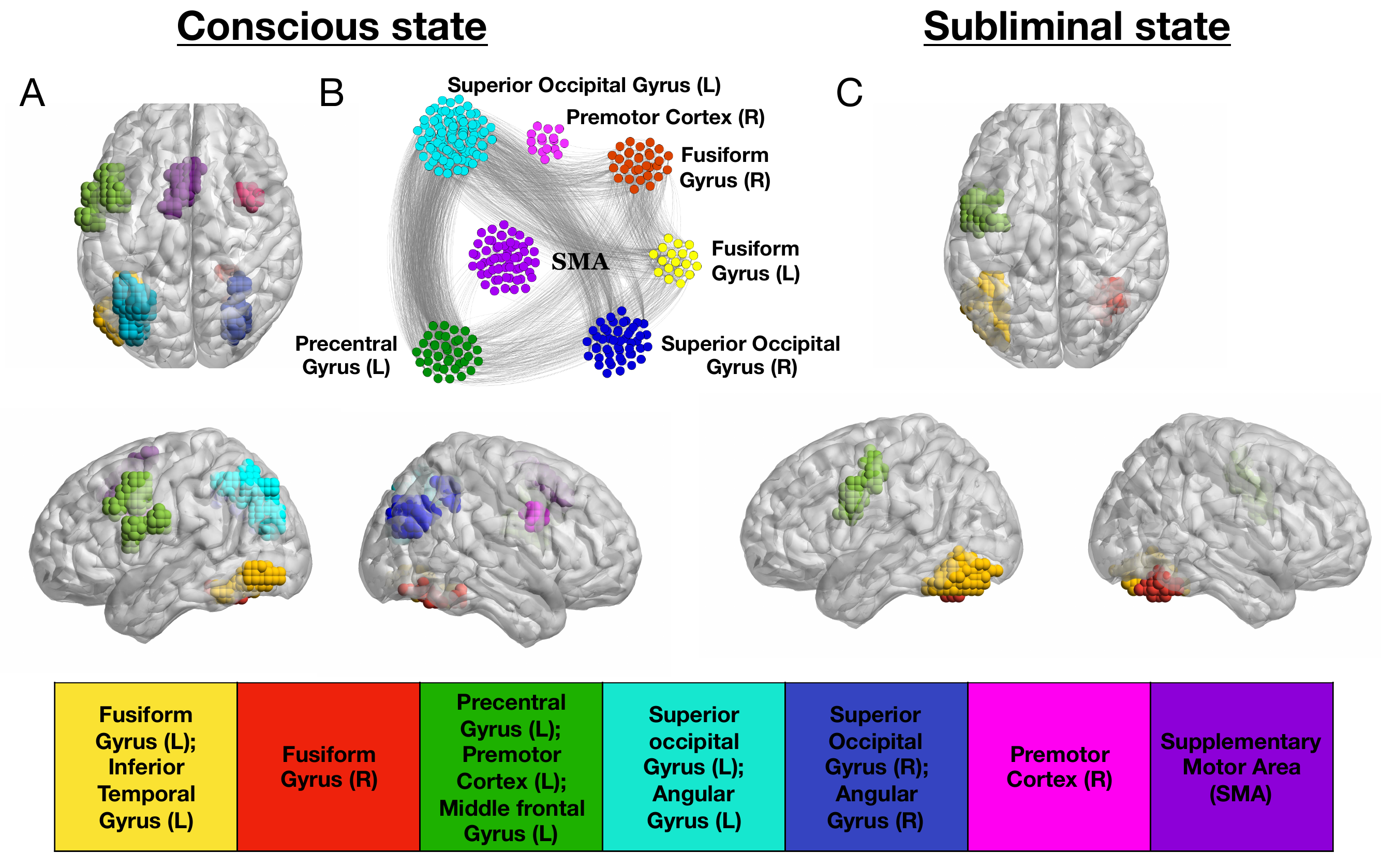}
\hfill
\caption{{\bf{Activation map and functional network of the conscious-state}}. a) Activation map of the conscious-state network for a representative subject ($P<10^{-6}$), left and right brain areas are abbreviated with L and R, respectively. Sagittal and axial view of the brain are shown. The fusiform gyrus and the inferior temporal lobe are involved in visual and word processing/recognition. The middle frontal gyrus is involved in working memory and attention, whereas the premotor cortex, the precentral gyrus, and the supplementary motor area (SMA) produce motor signaling. The superior occipital and the angular gyrus are involved in visual functions and transfer of visual information. b) Resulting functional network relative to the activation map of panel a) constructed with the procedure described in `Functional brain networks of the conscious state' Section. c) Activation map of the subliminal-state network ($P<10^{-2}$), $P$-values are chosen accordingly to Ref. \citep{experiment} in all panels.}
\label{AM_vis}
\end{figure*}
For each of these two states a
corresponding control sequence is presented in which the progression of
the images on the screen remains the same for each case, but blank screens are displayed 
instead of words (see Fig. \ref{experiment}). These control
conditions are used to estimate the background brain activity in
order to better evaluate the activation of each brain voxel \citep{experiment}. 
The fMRI signal was obtained for 15 subjects, and each of them repeated the experiment 5 times, thus 
data in \cite{experiment} it was collected for 75 different fMRI streams.

The acquired BOLD signal is processed using
SPM99 \citep{spmurl}. The fMRI time series are then analyzed by using the
widely adopted \textit{Generalized Linear model} \citep{GLM}, which produces as output the \textit{activation map} (AM). Fig. \ref{AM_vis}a shows the AM of the conscious state ($P<10^{-6}$), for a representative subject, whereas Fig. \ref{AM_vis}c illustrates the fMRI activation map of the subliminal state ($P<10^{-2}$) averaged across subjects  ($P$-values in both cases are chosen accordingly to \citep{experiment} and, for illustration, all the active nodes are shown with the same activation). Voxels are classified as belonging to a certain brain area with respect to their anatomical location and each brain area (module) is colored differently in the figures. For brevity, in the following, we will refer to all the active anatomical regions depicted with the same color in Fig. \ref{AM_vis} with the name of the first cluster in the legend, for each specific color.

By comparing the activation map of the {\it conscious} and {\it subliminal} state, shown respectively in Fig. \ref{AM_vis}a and Fig. \ref{AM_vis}c, we note that some brain regions are active in both brain states, i.e. the fusiform gyrus (yellow and red module) and the left precentral gyrus (green module). These regions are, in addition, the only ones characterizing the activation of the subliminal state. The fMRI activation in each conscious experiment spreads further and involves additional clusters, for the case shown in Fig. \ref{AM_vis}a, for instance, it involves the
left and right superior occipital gyrus (light blue and blue cluster respectively), the right
premotor cortex (pink cluster), and the supplementary motor area (SMA) (purple cluster).

\section*{Experimental Procedure}\label{sec:methods}

This Section illustrates the procedure we employ to investigate the transition from the conscious to the subliminal state, as described in  the Introduction Section. From the fMRI activation map of the scans acquired during the conscious-state experiment we construct the functional brain network of this brain state, for each individual and for each single stream.  Active fMRI voxels constitute the nodes of the 
network and links among these nodes are assigned by using pairwise correlations between the fMRI time series of the active voxels, with a procedure that we describe next (Fig. \ref{AM_vis}b shows one instance of this conscious network, corresponding to the fMRI activation of Fig. \ref{AM_vis}a).

Ideally, we would aim to build similar functional brain networks for the subliminal state, by employing the same procedure, so to have brain networks for the conscious- and subliminal-state experiment and then study the transition from one to the other. As discussed in Section Data  though, the subliminal state is characterized by a weaker overall fMRI activation, which therefore requires to apply a higher $P$-value threshold. As a consequence, noise effects play a greater role on the fMRI time series of the subliminal-state compared to the conscious-state time signals. Furthermore, the choice of a higher $P$-value produces subliminal AMs (obtained by comparing this state to the relative control stream) which present, in addition to clustered activity, isolated voxels spread across the brain that we consider false positives. In order to reduce these noise effects and get rid of this false positive activity, we averaged the activation maps of the subliminal state across streams and subjects, obtaining one final activation map for this state where voxels are active if appearing in at least 80\% of all activation maps. 
Furthermore, building a network for the subliminal state would not give any relevant information needed to explain the conscious $\to$ subliminal transition since the subliminal state is needed just to compare the reduction of the conscious network to validate any possible observation.
Therefore, we limited the use of the subliminal activation map as a benchmark for the study of the conscious $\to$ subliminal transition (Fig. \ref{AM_vis}c shows this final map) without constructing a subliminal-state functional network. 

After we have built the functional brain networks for the conscious-state we trim each of these structures by performing removal of the nodes belonging to different $k$-shells (peripherical nodes defined later in Sec. $k$-Core Percolation), from low to high value of $k$. The purpose of this trimming process it to identify those nodes which belong to each specific $k$-core and investigate whether, from this analysis, we can identify some pattern in the brain network structure of the conscious state, across subjects. We then investigate whether there exist markers which can help us identify the network differences between the conscious- and the subliminal-state as well as illuminate on the transition between these brain states. This procedure is described in details in Subsection $k$-core percolation.
%
%
%
%
\begin{figure}[t!]
\centering
\includegraphics[width=0.45\textwidth]{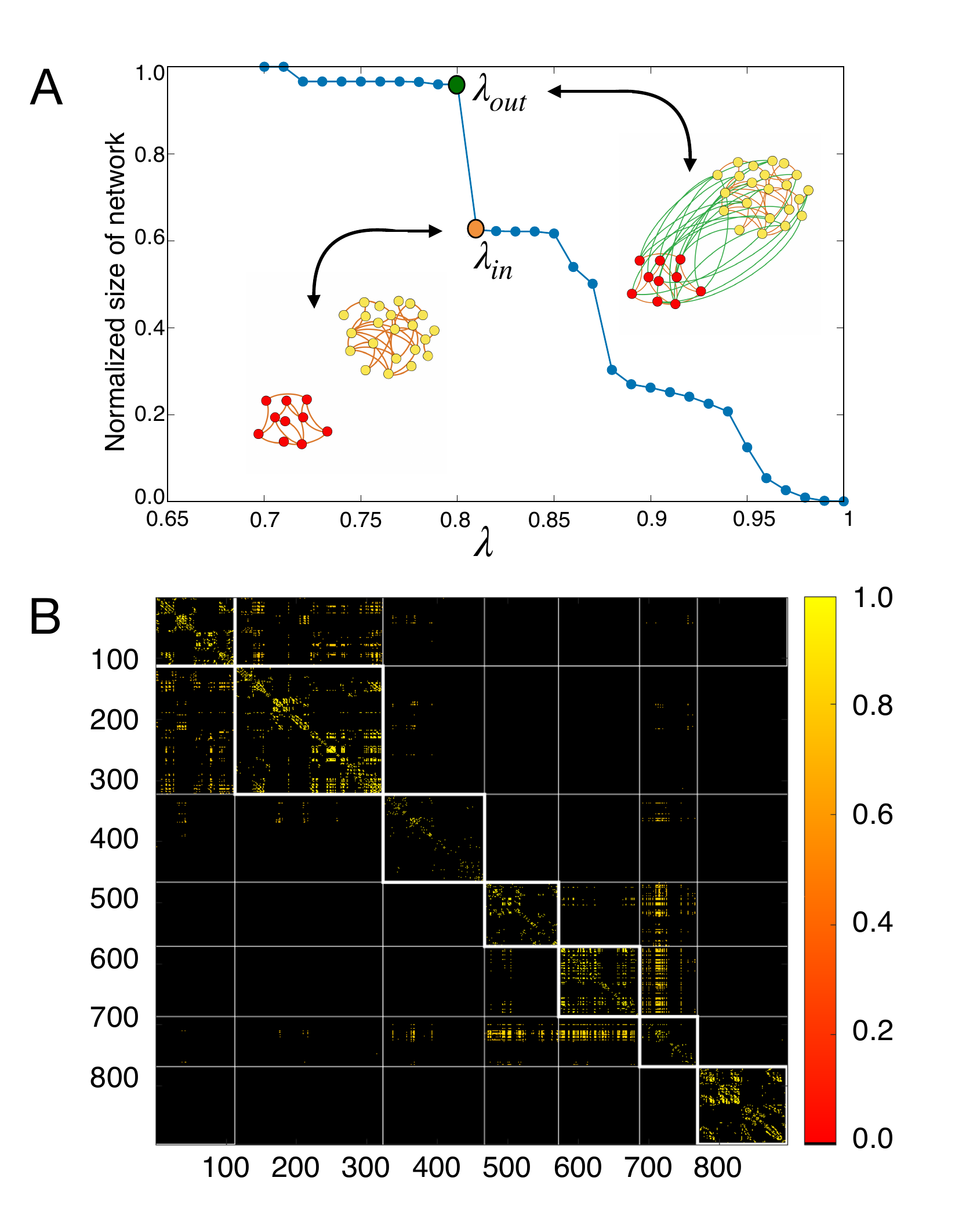}
\caption{ {\bf{Functional network construction}}. a) Percolation plot, i.e. the GCC of a network defined by the correlation matrix $C_{ij}(\lambda)$ vs the penalization parameter $\lambda$ is shown. The orange dot in the plot indicates the value of $\lambda_{in}$ used to fix the in-links within the two brain modules shown pictorially in the panel. The green dot pictures the value of $\lambda_{out}$ employed to fix the out-links connecting the same two modules together. b) Resulting thresholded correlation matrix according to panel a) of the functional network obtained with the above procedure.}
\label{perc_brain}
\end{figure}

\subsubsection*{Functional brain networks of the conscious state}\label{sec:functional}

For each subject and each fMRI stream of the conscious-state experiment we construct the functional brain network, following the approach described in  \citep{bullmore_sparse,gallos_percolation}, for a total of 75 networks (15 subjects, 5 fMRI streams each). 
As mentioned, the nodes in each of these networks are the active voxels in the corresponding fMRI activation map. Links are assigned based on the thresholded pairwise correlations of the registered fMRI signal between node $i$ and $j$, denoted as $C_{ij}$, as we explain in more details in the following. 

Brain networks show a modular organization \citep{sporns2013} where different brain regions are specialized in the performance of different cognitive tasks. In order to depict this modular structure, we group the active voxels in brain clusters according to their anatomical location (see Fig. \ref{AM_vis}). This spatial organization suggests to distinguish between links that connect nodes within the same cluster, that we call {\it in-links}, and long range edges connecting nodes in different clusters, that we call {\it out-links}, as described in \citep{gallos_percolation}.

Following standard literature \citep{bullmore_sparse,gallos_percolation}, we assign the links by thresholding the cross-correlation matrix in order to get rid of the weakest connections, such that two nodes $i$ and $j$ are wired together with the assigned weighted link $C_{ij}$ iff $C_{ij} \geq \lambda$, with $\lambda$ a tunable threshold parameter. Accordingly, each $\lambda$ threshold value defines a different functional network, identified by the thresholded correlation matrix that we denote as $C_{ij}(\lambda)$. The threshold parameter tunes the sparsity of this resulting network and therefore the size of its giant connected component (GCC). For $\lambda = 1$ the threshold is maximum and the nodes are isolated, so the GCC $= 0$. By decreasing $\lambda$, more and more nodes connect together and the GCC of the resulting network increases continuously until it reaches the unit value. Fig. \ref{perc_brain}a shows the behaviour of GCC vs $\lambda$ for a representative subject.

By following \citep{gallos_percolation}, we fix the $\lambda$ parameter through a `percolation' procedure described next.  Each sharp discontinuity in Fig. \ref{perc_brain}a
is due to the merging of connected brain clusters which abruptly increase the size of the GCC. At each one of these transitions we assign the {\it in-} and {\it out-links} of the brain clusters which connect together. An in-link between a node $i$ and $j$ within each of these clusters is assigned by $C_{ij}$ iff $C_{ij} \geq \lambda_{in}$, where $\lambda_{in}$ is the value of $\lambda$ right before the sharp transition at which the clusters merge occurs, coming from higher to lower $\lambda$ values (see orange dot in Fig. \ref{perc_brain}a). An out-link between a node $i$ and $j$ belonging to different clusters is assigned by the thresholded correlation matrix $C_{ij}(\lambda_{out})$, where $\lambda_{out}$ is the value of $\lambda$ right after the sharp transition (green dot in Fig. \ref{perc_brain}a). Fig. \ref{perc_brain}a illustrates this procedure and the resulting network pictorially, for a representative subject. The adjacency matrix of the final  architecture is shown in  Fig. \ref{perc_brain}b, where nodes are ordered sequentially according to their cluster association, in order to show the brain network's modular structure. 

We note that other methods \citep{funcnet_bassett,functnet_anzellotti} could be used to build the functional networks  . Our choice on the use of the percolation procedure to build such architecture was driven by the constraint that brain networks are sparse and such procedure guarantees sparsity by not overestimating the number of links between different clusters.
We stress that the wiring only reflects functional relations among fMRI active voxels and, in general, differs from the structural wiring obtained, for instance, through diffusion tensor imaging or other method of physical connectivity ~\citep{sporns_connectivity}. Thus, all our analysis and results must be considered as grounded on the functional network framework.
\begin{figure}[t!]
\centering
\includegraphics[width=.3\textwidth]{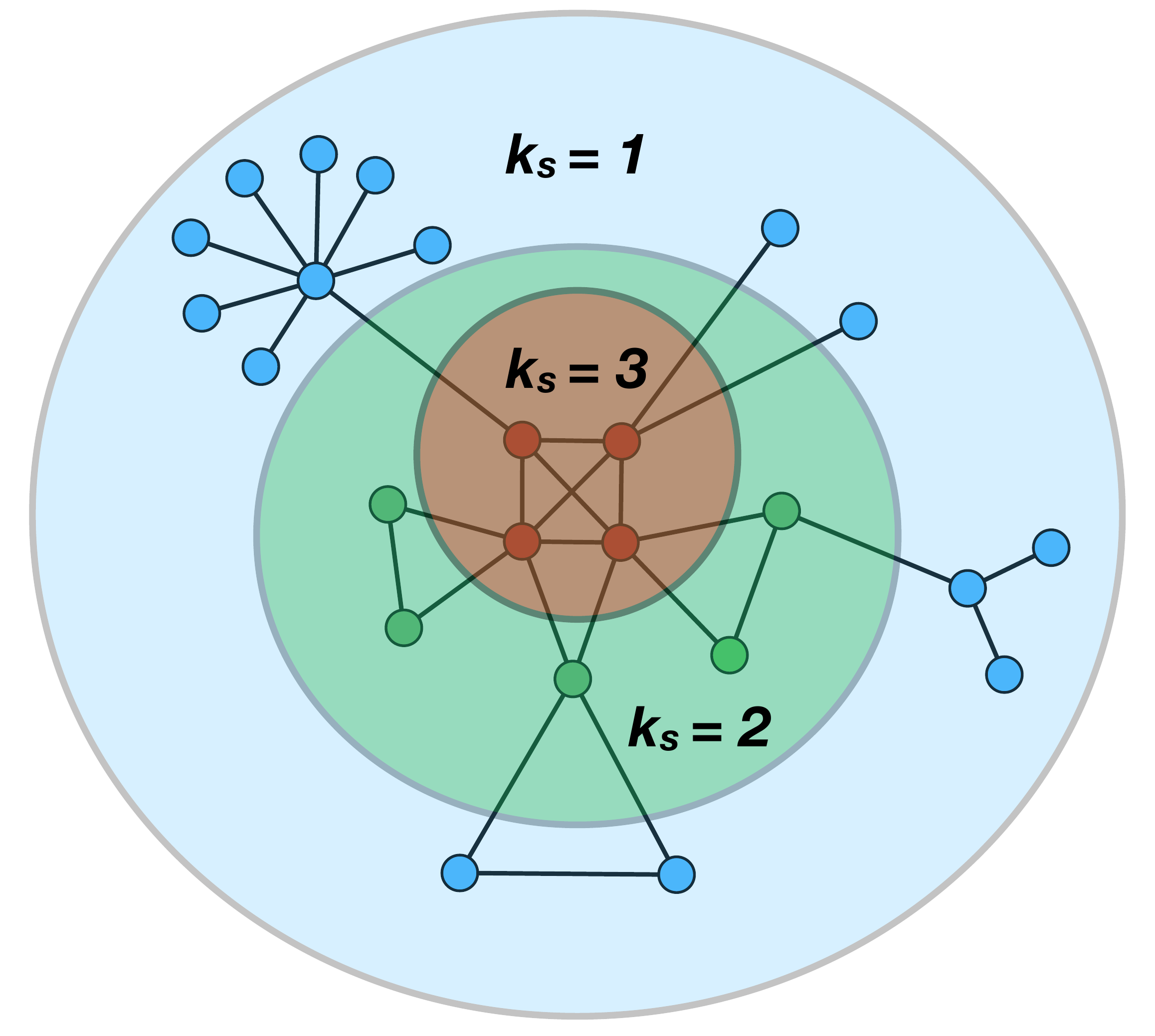}
\caption{ {\bf{Cartoon of a 3-core network.}} $k$-shells are defined as the set of nodes that belong to the $k$-core but not the $k+1$-core. As illustrated, $k$-shells are concentric; low $k$-shells are located in the outer part of the network, while for increasing $k$, nodes are situated in the most focal part until one reaches the highest $k$-shell, which corresponds to the maximum $k$-core, located at the center of the graph.}
\label{russian_doll}
\end{figure}
\subsubsection*{$k$-Core percolation} \label{kcore_percolation}

The concept of $k$-core has been firstly introduced in social sciences~\citep{intro_kcore} to describe network cohesion and, since then, it has been applied in many contexts, to describe robustness of random networks~\citep{dorogovtsev_kcore}, viral spreading in social networks ~\citep{kitsak} and large-scale structure of the brain ~\citep{sporns_kcore}.

For a given architecture, the $k$-core is the maximal subgraph, not necessarily globally connected, which consists of all the nodes with at least $k$ neighbours. This subnetwork can be obtained by removing iteratively all the nodes which have less than $k$ connections. Thus, to extract the $k$-core one starts pruning all the nodes with degree less than $k$. The removal of these nodes reduces the degree of their neighbors that can then drop below $k$. Thus, these nodes should be removed in turn, and the procedure iterates until no more nodes can be removed. The remaining structure is the $k$-core of the network (see Fig. \ref{russian_doll}).

For a given $k$, the $k$-core includes cores with higher $k$, thus the 1-core includes the 2-core, the 2-core includes the 3-core and so forth. Each $k$-core consists of the nodes in the periphery which is called $k$-shell (labelled $k_s$) and the resting $k+1$-core. The $k$-shell is, therefore, the region of the $k$-core which is not included in the $k+1$-core (see Fig. \ref{russian_doll}). Therefore the network has a nested structure made of $k$-core subnetworks with increasing $k$ and $k$-shells of order $k_s$. The innermost core of the network corresponds to the structure with the maximum $k$-core, called $k_{\rm core}^{\rm max}$, which is a topological invariant of the network \citep{dorogovtsev_kcore}.

For each one of the 75 conscious-state network, we can then compute the $k$-core and $k$-shell occupancy, i.e. the number of nodes which occupy each $k$-core or $k$-shell with a given $k$.  An interesting property of these networks emerges when one examines the nodes and the nodes' brain anatomical area in each $k$-shell. We discuss the results of this analysis in the next session.

\begin{figure}[t!]
\centering
\includegraphics[width=.45\textwidth]{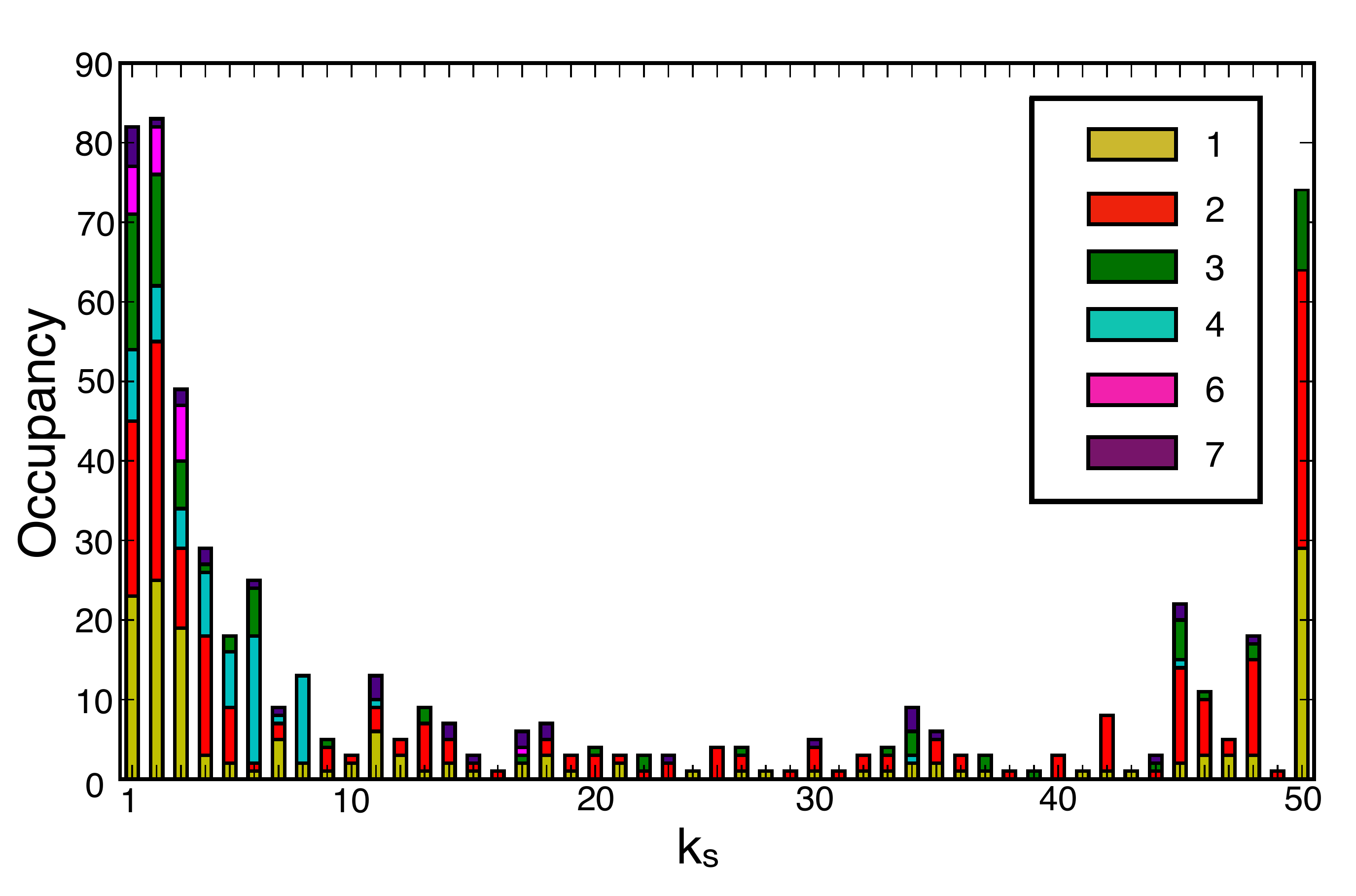}
 \caption{ {\bf{$k$-Shell occupancy for the conscious network}} of a representative subject.  The distribution presents a U-shape: high population of nodes in the lowest and highest $k$-shells. We observe that shells with the lowest $k$ are inhabited by nodes which belong to all the 7 brain clusters which are fMRI active in the brain. On the contrary, the maximum $k$-shell, the inner core of the network, is made by nodes which belong to only 3 clusters which, more importantly, are the only fMRI active clusters of the subliminal-state. }
 \label{occ_mod}
\end{figure}
\begin{figure*}[t!]
\centering
\includegraphics[width=0.9\textwidth]{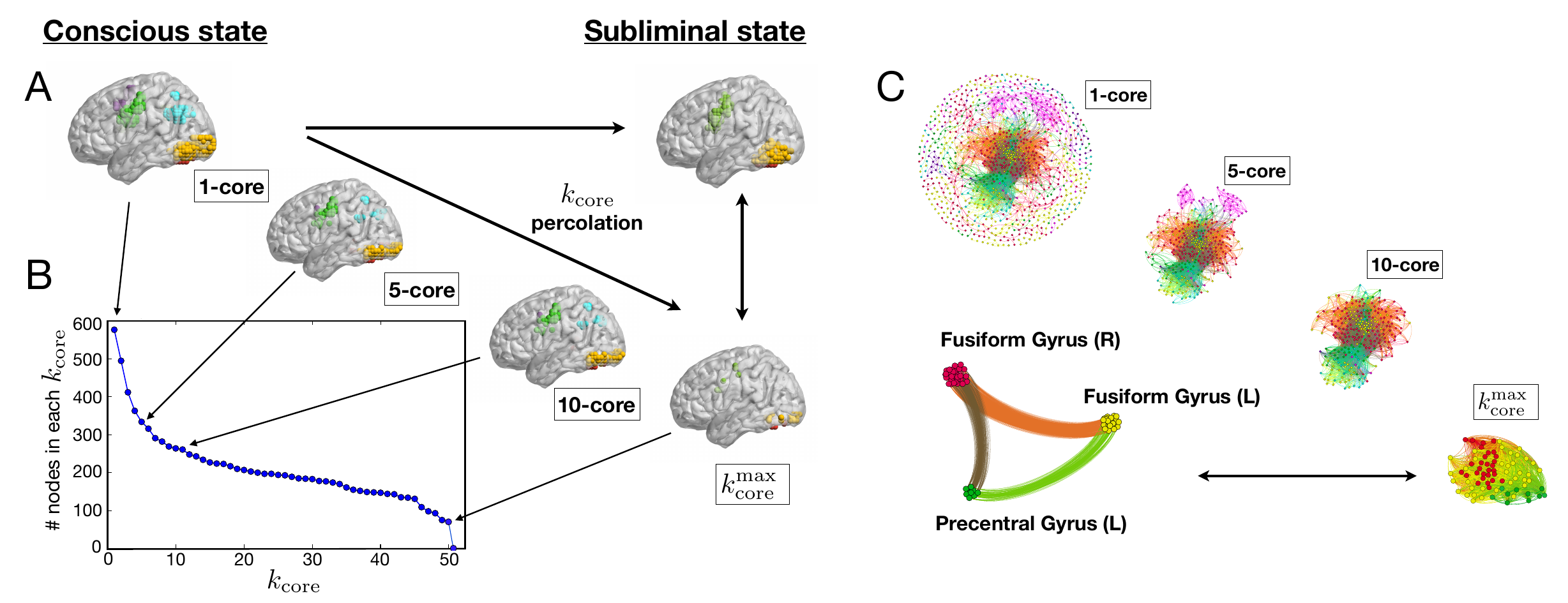} \hfill
\caption{{\bf{$k$-Core decomposition}} illustrated both in the activation brain map and in the functional network. a) Effect of the $k$-core trimming process on the activation map of the conscious-state, for increasing $k$-values. Nodes located in the low $k$-cores belong to all the brain clusters, on the contrary, nodes in the $k_{\rm core}^{\rm max}$ belong to the fusiform gyrus and left precentral gyrus (yellow, red, and green modules). b) Number of nodes in each $k$-core c) Same $k$-core decomposition of panel a), with a different visualization made on the functional network. Same considerations on the $k_{\rm core}^{\rm max}$ apply.}
\label{decomp}
\end{figure*} 
\section*{Results}\label{sec:results}

\subsubsection*{The maximal $k$-core of the conscious network corresponds to the
  subliminal-state}
  
Once we have performed the $k$-core trimming process as described in the previous Section, we can calculate the occupancy of each $k$-shell, for each subject, that is, the number of nodes in each $k$-shell. Fig. \ref{occ_mod} illustrates this occupancy for a representative conscious-state network. 
We note that, interestingly, the distribution presents a U-shape: it shows very high occupancy values for both very small and very high $k$-shell values, and low occupancy for the intermediate $k$-shell values. This shape of the distribution is consistent throughout all the conscious networks analyzed. Reference ~\cite{kate} attributes 
the U-shape of the occupancy distribution to the stability of the system: the high population of nodes in the lowest and highest 
$k$-shells suggests network robustness against both random local and global failure, thus making the brain 
a resilient system under these kind of perturbations. The same feature is also observed in ecosystems and financial networks ~\citep{kate} and it is general feature of many networks called the \textit{core-periphery structure} ~\citep{corradino_Ushape,krugman_Ushape,borgatti_Ushape,zhang_Ushape,verma_Ushape}.

We further observe that for the subject shown in Fig. \ref{occ_mod} the $k$-shells with small $k$'s are populated by nodes that belong to any fMRI active module and, therefore, nodes that are spread across all the active brain regions. More interestingly, Fig. \ref{occ_mod}
emphasizes that nodes which inhabit the $k$-shells with the highest  $k$ ($k=50$ for this subject) belong exclusively to those brain modules that are the only active clusters in the subliminal-state, namely the fusiform gyrus and the left precentral gyrus (yellow, red and green in Fig. \ref{AM_vis}c).
  
Figure \ref{decomp} shows pictorially the $k$-core decomposition process for the same subject presented in Fig. \ref{occ_mod}. Figure \ref{decomp}a illustrates the progressive inactivation of the fMRI active voxels, based on their $k$-core in the conscious-state network, while Fig. \ref{decomp}b shows the occupancy number in each of these $k$-cores, i.e. the number of nodes in each $k$-core. Figure \ref{decomp}c shows the $k$-core decomposition process by highlighting the network connectivity. As discussed for Fig. \ref{occ_mod}, Fig. \ref{decomp}c shows that the $k_{\rm core}^{\rm max}$ for this particular conscious-state network is made by the fusiform gyrus (yellow and red cluster) and by the left precentral gyrus (green cluster), which are the only fMRI active clusters in the averaged subliminal-state activation map.

\begin{figure}[t!]
\centering
\includegraphics[width=0.4\textwidth]{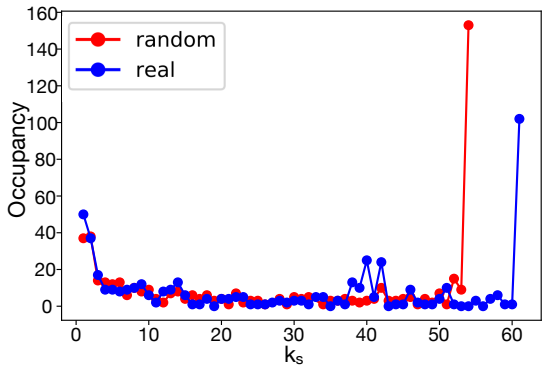}
\caption{{\bf{$k$-Shell occupancy distribution}} for a representative subject (all other subjects show similar results). Comparison between real conscious-state and a random control network generated as described in Section Results. Random networks exhibit a lower value of the maximum $k$-shells compared to the real network, a behavior common to all the conscious-state networks ($p<10^{-6}$).}
\label{ks_rand}
\end{figure}

To verify whether these results are statistically significant we perform the following analysis. For each of the 75 conscious-state graphs we generated $10^6$ new architectures obtained by randomly rewiring the original network and keeping constant the degree of each node. We then apply $k$-core decomposition to each of these $10^6$ random networks and compute the $k$-shell occupancy. The obtained occupancy is then averaged across the generated random networks and compared with the $k$-shell occupancy of the conscious-state network from which the random architectures are created. Results are shown in Fig. \ref{ks_rand} for a representative subject and illustrate two interesting points. First, the functional conscious-state network has a much higher $k_{\rm core}^{\rm max}$ than the averaged random case. Second, the average occupancy distribution of the random networks shows the same U-shape feature that we found in all the functional conscious-state networks obtained from the fMRI signal. This suggests that the source of this shape is related to the degree distribution of the nodes, being this distribution the same both in real conscious-state and randomly generated networks. We mention that although Fig. \ref{ks_rand} refers to a representative case, we found qualitatively the same results for all the 75 conscious-state networks.

\begin{figure}[t!]
\centering
\includegraphics[width=0.5\textwidth]{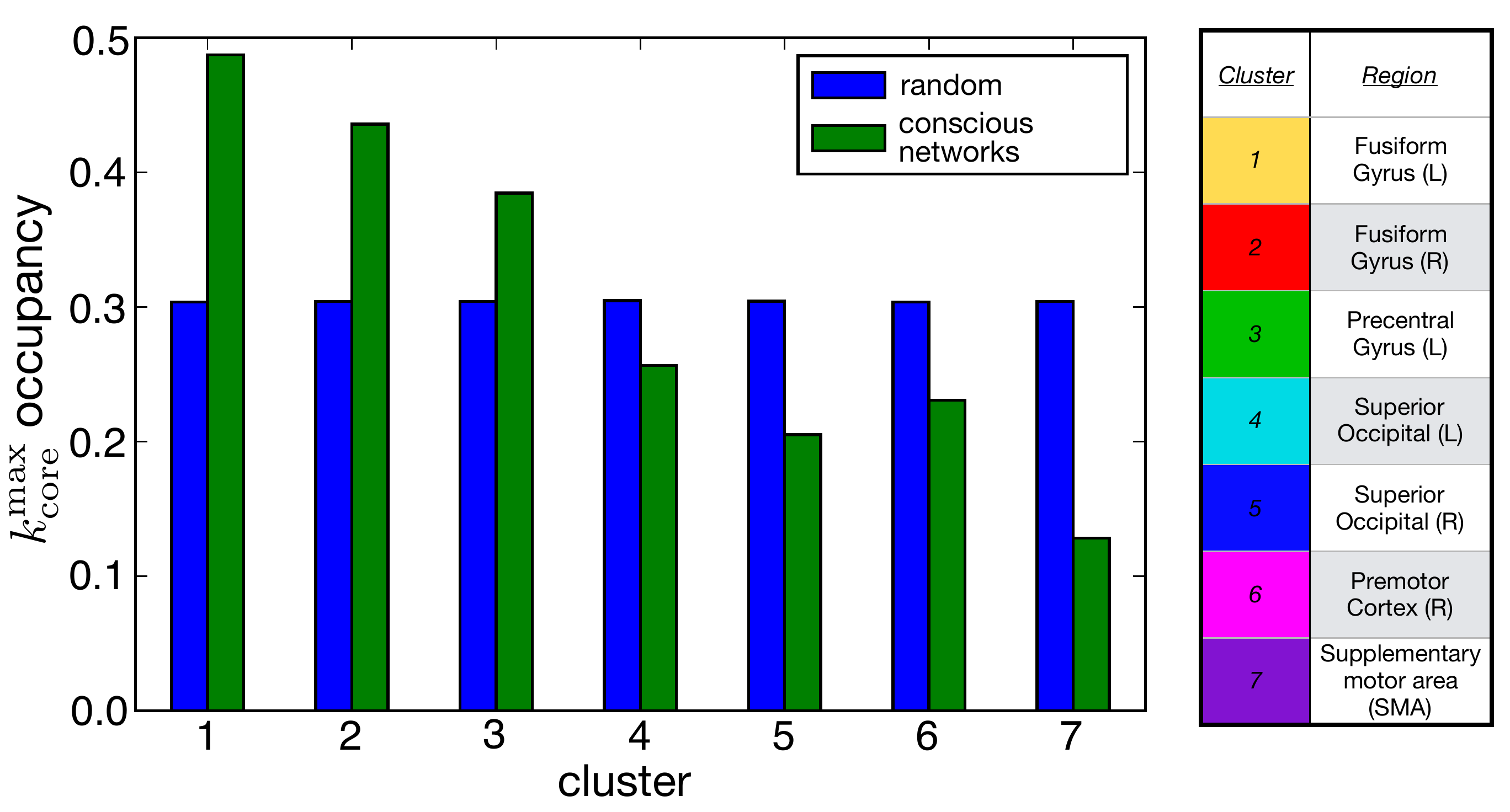}
\caption{{\bf Cluster occupancy of nodes in the maximum $k$-core} (which coincide with the maximum $k$-shell). Green bars show the normalized cluster occupancy of nodes which populate the maximum $k$-core. If the occupation of the $k$-core were due to a random effect then one would find a distribution of about 30\% in each cluster (blue bars). We observe that the left and right fusiform gyrus and the left prefrontal gyrus populate the maximum $k$-core more than what is expected at random ($P<10^{-5}$).}
\label{occ_mod2}
\end{figure}

As already noted, results of Fig. \ref{decomp} illustrate that, for this representative subject, nodes which populate the $k_{\rm core}^{\rm max}$ of the conscious-state network belong to those clusters which are the only active ones in the averaged subliminal-state activation map (see Fig. \ref{AM_vis}c). In order to check whether this result is consistent across all the conscious-state networks we performed a group analysis at the clusters level. For each of the 75 conscious networks we assign count 1 to the cluster of nodes belonging to the  $k_{\rm core}^{\rm max}$. For instance, for the particular network of Fig. \ref{occ_mod} we assign count 1 to clusters 1, 2 and 3 (yellow, red and green respectively, which populate the $k_{\rm core}^{\rm max} = 50$). The normalized occupation number of the clusters in the $k_{\rm core}^{\rm max}$ across conscious-state networks is shown in Fig. \ref{occ_mod2}, where clusters have been ordered progressively. The green histogram represents the frequency with which each cluster appears in the $k_{\rm core}^{\rm max}$. These results are compared to random ones illustrated by the blue histogram. The probability that each cluster is, at random, in the $k_{\rm core}^{\rm max}$ is $1/7$ of the sum of counts of all modules which, when normalized with respect to the number of conscious networks analyzed (75 networks), translates to a 30\% chance to populate the $k_{\rm core}^{\rm max}$ (see Fig. \ref{occ_mod2}).

This comparison not only shows that the fusiform gyrus and the left precentral gyrus (cluster 1, 2 and 3 in Fig. \ref{occ_mod2}) are those which mostly populate the $k_{\rm core}^{\rm max}$ of the conscious-state network across subjects, it also points out that this is not due to a random effect ($P =10^{-5}$). As reported above, these clusters are the only active ones in the subliminal-state experiment. The other four clusters (4 to 7 in Fig. \ref{occ_mod2}) populate the $k_{\rm core}^{\rm max}$ with a normalized frequency which is less than a random effect, making  their presence in the maximum $k$-core statistically less significant. 

\begin{figure}[t!]
\centering
\includegraphics[width=0.5\textwidth]{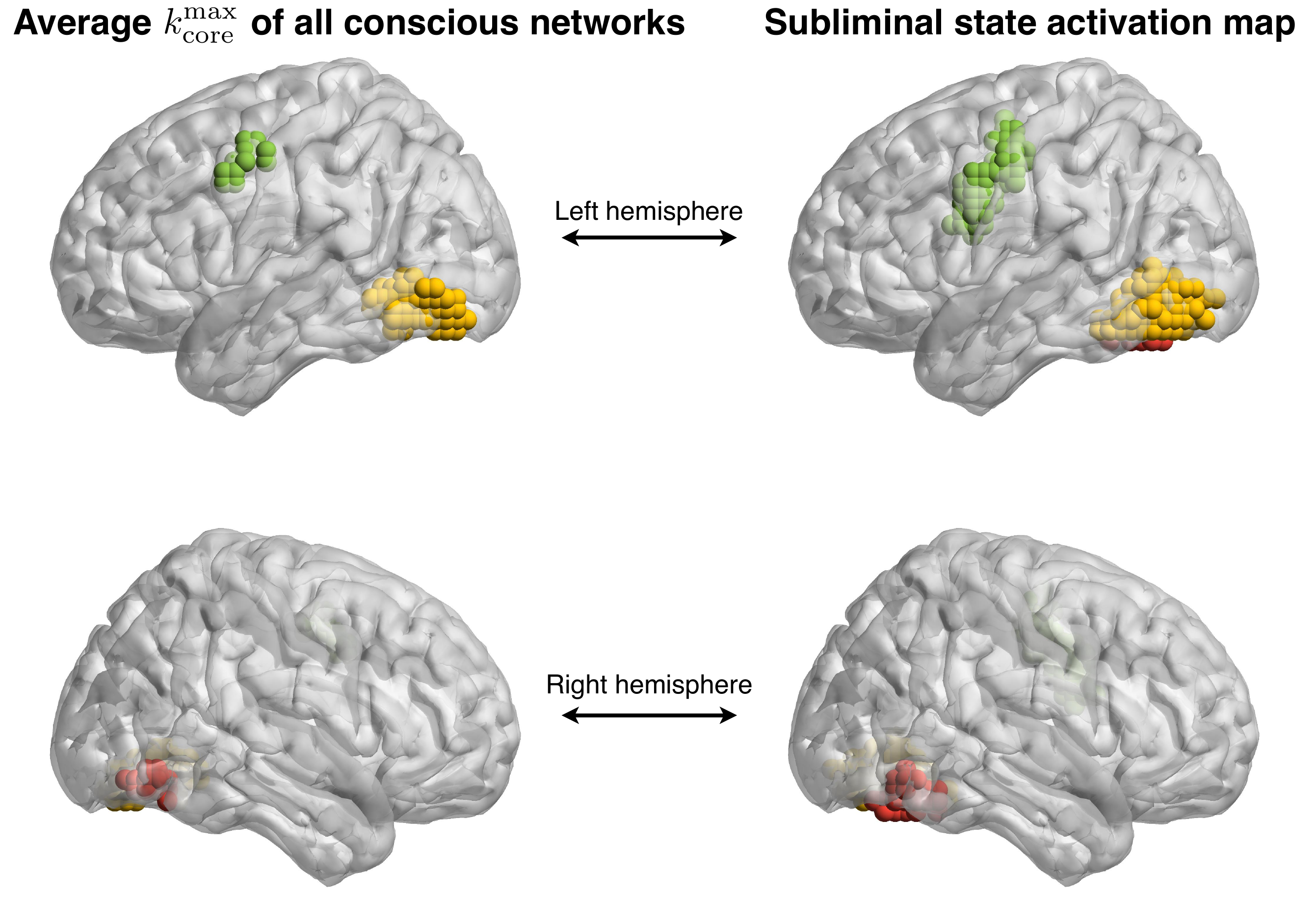}
\caption{{\bf{Statistical analysis at the node level}}. Left panels: nodes which are in the $k_{\rm core}^{\rm max}$ of all conscious-state networks that pass the t-test  ($p<10^{-5}$) when compared  with the subliminal-state. Right panel: activation map of the subliminal-state shown for comparison. Roughly 1/3 of the nodes which belong to the $k_{\rm core}^{\rm max}$ of the conscious-state also belong to the subliminal-state activation map. The subliminal state (right panel) largely overlaps with the maximum $k$-core of the conscious network (left panel).}
\label{conc}
\end{figure}

As a further test, we check whether the above group results at the cluster level are also consistent at the node level. In other words, we investigate whether, across networks, the nodes in the $k_{\rm core}^{\rm max}$ of the conscious-state are the same nodes which are active in the subliminal-state. For each conscious-state network we compute how many nodes ($n_k$) are in the $k_{\rm core}^{\rm max}$, and we check how many of these nodes are also in the activation map of the subliminal-state, we refer to this number as $n_x$. Then, in each conscious-state network we randomly select $n_k$ nodes and check how many of them also belong to the activation map of the subliminal-state. We repeat this random sampling $10^5$ times in order to have a distribution of randomly selected nodes in the conscious network which are also in the subliminal activation map and perform a $t$-test with the overlap $n_x$ described above. In Fig. \ref{conc} we show the results for those networks which passed the $t$-test. In details, the left panels show nodes which are both part of the $k_{\rm core}^{\rm max}$ of the conscious-state and of the AM of the subliminal-state (show in the right panels for comparison), consistently across networks. Quantitatively, nearly $1/3$ of the nodes in the AM of the subliminal-state also belong  to the $k_{\rm core}^{\rm max}$ of all the conscious functional networks that pass the $t$-test (precisely, 112 nodes over 340). This suggest that the subliminal-state, which remains active during period of non-conscious perception, constitutes a large part of the $k_{\rm core}^{\rm max}$ of the conscious state.

\section*{Discussion}\label{sec:discussion}

In this Section we elaborate on the interpretation of the results by introducing a dynamical model describing the time evolution of mutualistic complex systems \citep{kcore} which directly addressed the stability of the network and its relation to the $k$-core. The model of ~\cite{kcore} applies to the complex networks with only positive interaction. Since the correlations between nodes of the conscious-state networks turn out to be positive, the dynamics of these brain networks can be modeled with differential equations accounting for mutualistic interplay proposed in Ref.~\cite{kcore}. Recent results \citep{kcore} show that, for such mutualistic systems, the $k_{\rm core}^{\rm max}$ of the network is the most resilient structure, i.e. the last architecture which collapses due to the weakening of the interactions strength. Findings discussed in Section Results show that the $k_{\rm core}^{\rm max}$ of the conscious-state largely overlap with the subliminal-state activation map. Thus, in short, the findings of \cite{kcore} help us interpreting the subliminal-state as the most resilient part of the conscious-state when the correlations (interactions) strength is weakened. 

This leads us to theorize that the conscious $\to$ subliminal transition in the brain happens through the weakening of the interactions strength in the network. In other words, regions which are highly correlated in the conscious-state suddenly become less correlated and, therefore, not fMRI active thus producing a subliminal state of activation. We see this effect in the data through the activation map shown in Fig. \ref{AM_vis} which shows that, indeed, the subliminal state is characterized by much less fMRI activation (compare Fig. \ref{AM_vis}a with Fig. \ref{AM_vis}c). Furthermore, findings discussed in Section Results demonstrate that the residual activation of the subliminal-state largely matches with the $k_{\rm core}^{\rm max}$ of the functional conscious-state network. These two evidences are consistent with the results discussed in \citep{kcore} which help us interpreting the transition conscious $\to$ subliminal as taking place through the activity collapse of certain conscious-state area, due to decrease covariation among these areas, leaving residually active only areas in the $k_{\rm core}^{\rm max}$. We conclude that these areas are therefore the most resilient ones, as shown in \cite{kcore}, and, in our case, those which characterize the subliminal brain activity. 

In the next section we briefly review the relevant results of Ref. \cite{kcore} to the above discussion in order to elaborate an interpretation of our findings and a description of the conscious $\to$ subliminal transition at the functional network level.

\subsubsection*{Dynamical model and $k$-core percolation} \label{model}

Dynamics of a neural network can be described by a model of coupled interacting neurons through sigmoidal responses  \citep{kcore_neural2,kcore_neural1}. Here, by coarse graining the neural activity, we use the same model  to describe the dynamical evolution of the fMRI signal in each voxel, with the following nonlinear differential equations \citep{kcore_neural2,kcore_neural1}

\begin{equation}
\dot{x_i}(t)= I - \frac{x_i}{R} + \frac{1}{2}\sum_{j=1}^{N}A_{ij} J_{ij}\big[1+{\rm tanh}(n(x_j-\alpha))\big]
\label{neuralnetworks}
\end{equation}
Here, $x_i(t)$ is the fMRI activity of voxel $i$, $N$ is the number of voxels, $I$ is the
background BOLD activity, $R$ is the inverse of the inactivation rate,
$n$ is the slope of the sigmoid function, $\alpha$ is a BOLD activity
threshold at the fMRI voxel level, $A_{ij}$ is the adjacency matrix, i.e. $A_{ij} = 1$ if voxel $i$ and $j$ are connected and zero otherwise and $J_{ij}$ is the interaction strength between pair of
voxels where we take $J_{ij}=C_{ij}$ as the strength of correlations from the data. Notice that the theory is only valid in absence of inhibition, i.e when $J_{ij}>0$. Let us note that  the matrix shown in Fig. \ref{perc_brain}b includes both the information encoded in $A_{ij}$ (whether a link is present or not) and in $J_{ij}$ (the strength of such link). In Eq. \eqref{neuralnetworks} we employ this slightly different formalism for consistency with Ref. \cite{kcore}. 

For a given set of initial conditions, the fixed point solution of Eq. \eqref{neuralnetworks} is completely determined by the values of the dynamical parameters. Of particular interest is the identification of the tipping point by tuning of these parameters, i.e. the point at which all nodes are inactive ($x_i = 0$ for each $i$). In general, the analytical derivation of the fixed point solution of Eq. \eqref{neuralnetworks} is too cumbersome. Morone {\it et al.} in \citep{kcore} have shown, yet, that under the assumptions of constant couplings ($J_{ij} = J$ for all $i,j$) and by replacing  $\frac{1}{2}\big[1+{\rm tanh}(n(x_j-\alpha))\big] \approx \Theta(x_j
 -\alpha)$, where $\Theta(x)$ is the Heaviside function, it is possible to obtain an approximate solution of this tipping point.  
 
We observe that, in the data of the present experiments \citep{experiment}, analyzed and discussed in the previous Sections, the
interactions among voxels are mainly positive (see Fig. \ref{perc_brain}b). This outcome could be explained by noting that the fMRI signal is stimulus-driven (words shown on a screen) and, therefore, the correlation coefficient among the fMRI activity of two voxels which follow the same stimulus is most likely positive. If we then assume that these interactions (correlations) have all the same strength $J$, by following the approximation of \cite{kcore} for the sigmoidal function ($n \to \infty$), we can write the steady-state solution of $x_i^*$ from Eq. \eqref{neuralnetworks} as:
\begin{equation}\label{eq:xJ}
x^\ast_i(t)= IR + JR \sum_{j=1}^{N} A_{ij}\Theta(x^\ast_j -\alpha)
\end{equation}
and, with the following change of variable
\begin{equation}
y^\ast_i=\frac{x^\ast_i - IR}{JR},
\end{equation}
Eq. \eqref{eq:xJ} can be written in terms of the reduced density $y^\ast$, in the following form: 

\begin{equation}
y^\ast_i(t)=\sum_{j=1}^{N} A_{ij}\Theta(y^\ast_j-K_J),
\label{ystarneural}
\end{equation}
with
\begin{equation}
K_J=\frac{\alpha}{JR}- \frac{I}{J}.
\label{eq:K}
\end{equation}
The parameter $K_J$ in Eq. \eqref{ystarneural} controls the threshold of mutualistic benefit \citep{kcore} which, in the case of study, is a threshold of mutualistic signal enhancement between voxels. In practice, $K_J$ is the threshold in the $\Theta$-function of Eq. \eqref{ystarneural} which allows voxel $i$ to increase its activation thanks to the interaction with voxel $j$ only when the densities $y_i^*$ are greater than $K_J$. Let us observe that $K_J$ is inversely proportional to the interaction strength $J$. By weakening the interactions the threshold increases and, thus, the final activity of voxel $i$ decreases. In other words, by keep decreasing the interactions, the activity of some of the voxels $y_j^\ast$ falls under the threshold and therefore confers no activation to $y_i^\ast$ (see Eq. \eqref{ystarneural}). Hence, from Eqs. \eqref{ystarneural} and \eqref{eq:K} it is clear that there exist a critical value $J_c$, and thus a critical threshold $K_J(J_c)$, at which the only solution of Eq. \eqref{ystarneural} is $y_i^\ast=0$ for all $i$. 

In Ref. \cite{kcore} the authors show that this critical threshold is related to the maximum $k$-core of the network. Indeed,  the reduced density $y_i^*$ assumes only integer values in the set $y_i^* \in \{1,\dots,k_i\}$, where $k_i$ is the degree of voxel $i$. For a given threshold $K_J$, voxels with degree $k_j < K_J$ do not contribute to Eq. \eqref{ystarneural}, so they can be removed from the network. After this removal, some of the remaining voxels will have a smaller degree $k_j'$ (due to the fact that they have lost some of their neighbors with the removal). Voxels with $k_j'<K_J$ can then be removed in turn, because they will not contribute to Eq. \eqref{ystarneural}, and so forth. This process is exactly the algorithm for extracting the $K_J$-core from a network and voxels remaining at the end of this procedure are the voxels belonging to the $K_J$-core \citep{kcore}. By increasing the threshold $K_J$ from low to higher values, voxels from the low-to-higher $k$-cores will cease to contribute to the dynamics of the network, until the critical threshold $K_J(J_c) = k_{\rm core}^{\rm max} $ is reached. Above this threshold, the only fixed point solution is the network collapse $ y_i^\ast = 0$ for all $i$.

From this findings it results that, as also mentioned at the beginning of the Discussion Section, the $ k_{\rm core}^{\rm max} $ structure is the most resilient part of a network to the decreasing of the interactions strength. Based on these findings, we interpret the conscious $\to$ subliminal transition as a passage from high to lower correlations among brain areas which ends in a final state, i.e. the subliminal, that corresponds to the $k_{\rm core}^{\rm max} $ of the conscious-state network. 
It is worth mentioning that this dynamical model could be applied to any experiment on consciousness that results in positive interactions between active nodes assuming that the underlying dynamics is the one described by Eq. \eqref{neuralnetworks}.

\subsubsection*{Our results in light of Libet and Dehaene's consciousness theories}

Two prominent theories of the relation between unconscious information and conscious access have been developed by Benjamin Libet and Stanislas Dehaene. The former stressed how, through the analysis of EEG data, all external stimuli is processed in the brain unconsciously a couple of hundred of milliseconds before any voluntary act. According to this theory, unconscious information is the spark for the initiation of all conscious actions, and there is a role for consciousness and executive control to regulate actions of information processed subliminally \citep{Libet}. On the other hand, Stanislas Dehaene, has shown the existence of a large-scale versatile brain system that involves regions in the parietal and frontal cortex that set a temporary workspace to bind and share information \citep{framework,book,monkeys}. This framework which allows exchange of information through first bottom up, followed by top down propagation, is referred to as ignition; if the incoming stimuli does not activate voxels strongly enough, then the information will not be manifested consciously by the brain.

Our findings add a new view which is consistent with these theories. Both share the notion that, while conscious activation involves the non-linear and massive activation of a broad set of brain areas in an ignition process, the onset of this mechanism is in local-circuits, which encode information for this specific process that might eventually become conscious. Our work shows that despite the massive propagation of information the core of activity, which is at the seed of the unconscious-state, remains at the deepest core, i.e. at the shell structure of the functional networks. This finding is quite reminiscent of the 'theory of vision' proposed by David Mumford and colleagues \citep{Lee}. This theory argues that V1 is a high frequency functional core of the brain. It buffers and holds temporarily (as in a blackboard, or as in a workspace) information for which its receptive fields are optimally suited. In other words, a core shell of conscious activation, may not be a common set of neurons but, instead, it may vary according to the functional requirements of the specific conscious percept at any given time.

\section*{Summary}\label{sec:conclusions}

In this work we investigated the conscious $\to$ subliminal transition in the brain through the network analysis of fMRI data collected in Ref. \cite{experiment} where two experiments on human subjects were performed, specifically designed to induce either a conscious or a non-conscious (subliminal) perception of a word flashed on a screen.  

From the data we first observed that fMRI activation of the subliminal-state is largely a subset of the activation of the conscious-state. Furthermore, we note that links in the functional brain network of the conscious-state, built from the fMRI signal, are mostly positive. 

These two observations from the data analysis, together with the recent findings of Ref. \cite{kcore}, led us to perform a $k$-core study of the conscious-state network structure. The authors of \cite{kcore}  recognized indeed that, under certain approximations, neural dynamical systems with positive interactions show a decrease of their activation to the weakening of their interactions strength. The most resilient part of the network to this kind of weakening, i.e. the last structure to remain active, is the $k_{\rm core}^{\rm max}$ of the system. So, driven by the above observations we investigated whether the subliminal-state was related to the $k_{\rm core}^{\rm max}$ of the conscious-state. 

We found that, at the cluster level, the subliminal-state is made of fMRI active clusters which are those that most populate the $k_{\rm core}^{\rm max}$ of the conscious-state network, across subjects and experiments. At the node level, we found that roughly $1/3$ of the active voxels of the subliminal-state exactly overlap (node by node) with the nodes in the $k_{\rm core}^{\rm max}$ of the conscious-state network, across fMRI streams. To verify that these results were not due to chance, we also compared them with outcomes obtained from suitable randomly generated models. 

Overall, these findings are in agreement with the prediction of Ref. \citep{kcore} and led us to conclude that the conscious $\to$ subliminal transition may be interpreted as caused by a decrease of the correlated fMRI activity among voxels, due to the fMRI inhibition of certain brain areas. The areas which survive this inhibition, i.e. those which constitute the subliminal-state, are also those that, statistically, belong to the most resilient structure of the conscious-state network: the $k_{\rm core}^{\rm max}$. This not only sheds light on the nature of the conscious $\to$ subliminal transition but, furthermore, motivates us to interpret the subliminal-state activity as the most robust to the weakening of the fMRI signal. Indeed, this state is the one which persists as background fMRI activity when non-conscious perception is present and the state from which conscious perception arises.

The conscious $\to$ subliminal transition is a profound and intriguing problem in neuroscience and this work certainly does not answer all the questions that it rises. On the other hand, from a system neuroscience perspective, we think our results highlight the importance of studying network structures that could unveil useful patterns or markers able to shed light on similarity and differences between conscious and subliminal awareness and on the transition from one to the other. \newline

{\bf Acknowledgements}. We thank Flaviano Morone and Lucas Parra for
useful discussions and insights and Luca Pasquini for help with brain anatomy. We thank Stanislas Dehaene for
graciously sharing his data from Ref. \cite{experiment} and allowing
us to analyze this dataset. This work is supported by NIH-NIGMS
R01EB022720, NIH-NCI U54CA137788 / \\ U54CA132378 and NSF-IIS 1515022.

\bibliographystyle{abbrvnat}
\bibliography{transition_bib2}

\end{document}